\documentclass[%
 reprint,
 amsmath,amssymb,
 aps,
]{revtex4-2}

\usepackage{graphicx}
\usepackage{dcolumn}
\usepackage{bm}

\begin{document}

\preprint{APS/123-QED}
		
		\widetext
		\leftline{submitted on \today}
		\leftline{Primary authors: Anil Raghav}
		\leftline{To be submitted to xx } 
		\leftline{Comment to {\tt raghavanil1984@gmail.com} }
		
		
		\title{The Solar Enigma: Decoding the Mystery of the Hot Corona and Solar Wind Acceleration}

		\author{Anil Narayan Raghav}
		\affiliation{University Department of Physics, University of Mumbai, Vidyanagari, Santacruz (E), Mumbai-400098, India}


		\date{\today}

		\begin{abstract}

The Sun’s corona, heated to temperatures exceeding one million Kelvin despite lying above the cooler photosphere, has puzzled astrophysicists since its discovery in the 1940s. Prevailing theories, based on acoustic and magnetohydrodynamic waves, and micro/nano-flares, fail to fully account for this extreme heating or the origins of the supersonic solar wind. We propose a mechanism in which magnetic reconnection rapidly reconfigures field lines and exerts a torque on the magnetic moment of gyrating charged particles. This torque transfers magnetic energy to the particles, effectively doubling their energy and heating the plasma. As particles traverse multiple reconnection sites, they gradually heat to coronal temperatures. With just a few additional crossings, particles can gain enough energy to overcome the Sun’s gravity, directly driving solar wind acceleration. Our hypothesis resolves the coronal heating paradox and unifies solar wind dynamics under a universal astrophysical process. This work redefines our understanding of energy transfer in magnetized plasmas, with implications extending to stellar and astrophysical systems.

\end{abstract}
\maketitle





\section{Introduction} \label{sec:intro}
The solar corona, the Sun’s tenuous outer atmosphere, appears as a pearly glow during total eclipses \cite{saito1973arch} and remains one of most enduring mysteries in astrophysics. Its temperature exceeds 1 million Kelvin, a phenomenon first inferred from spectroscopic observations in the 1940s \cite{edlen1937kenntnis,grotian1939frage,lyot1939study,edlen1945identification}. This extreme heating, nearly 300 times hotter than the underlying photosphere, defies classical thermodynamics, as the corona’s low density prevents significant conductive heat transfer from the Sun’s surface \cite{sakurai2017heating}.

Despite decades of research, the precise mechanism responsible for coronal heating remains unresolved. Advances in high-resolution observations, magnetohydrodynamic (MHD) simulations, and theoretical modeling have refined our understanding, yet proposed mechanisms, including alternating current (wave-driven) and direct current (magnetic reconnection) processes, fail to fully account for the observed energy budget and plasma dynamics (see review articles \cite{cranmer2019properties,billings1966guide,withbroe1977mass,kuperus1981theory,narain1990chromospheric,low1996solar,aschwanden2006physics,klimchuk2006solving,golub2010solar,parnell2012contemporary,reale2014coronal,schmelz2015can,velli2015models,shimizu2018first}). Resolving this enigma is critical for understanding solar wind acceleration and magnetically coupled processes throughout the heliosphere and astrophysical plasma.

The first theoretical framework for coronal heating, proposed in 1948, attributed it to acoustic waves generated by turbulent convection in the photosphere \cite{biermann1948ursache}. Pressure perturbations from granulation-driven motions were thought to propagate upward, amplifying as they passed through the chromosphere’s steep density gradient. In the low-density corona, nonlinear wave steepening was expected to produce shock waves that dissipate energy and sustain high temperatures \cite{schwarzschild1948noise}. However, later analyses showed that the available acoustic wave energy flux was orders of magnitude too low to counteract radiative losses \cite{lighthill1954sound,unno1966generation,athay1978chromospheric,cranmer2007self}. This realization, along with the corona’s strong magnetic field, shifted focus toward magnetically driven heating mechanisms.

The corona’s low plasma beta ($\beta \ll 1$) suggests that magnetic forces dominate energy transport, unlike in the photosphere ($\beta > 1$), where thermal pressure plays a larger role. MHD wave-driven models propose that Alfvén, fast-mode, and slow-mode waves, generated at the solar surface, transport energy into the corona \cite{osterbrock1961heating}. Slow-mode waves dissipate in the chromosphere, similar to acoustic damping, while fast-mode waves require nonlinear steepening to form shocks, an inefficient process in the corona’s weak density gradients. Alfvén waves, central to alternating current heating models, can dissipate via resonant absorption \cite{davila1987heating}, phase mixing \cite{heyvaerts1983coronal,sakurai1984generation}, or turbulent cascades from nonlinear interactions \cite{chin1972nonlinear}. While these mechanisms align with broader magnetic energy conversion theories, their efficiency in sustaining million-Kelvin temperatures remains an open question.

Observational constraints further challenge acoustic heating models. Measurements at the chromosphere’s upper boundary show an acoustic wave flux of only $10^4$ erg cm$^{-2}$ s$^{-1}$, three orders of magnitude below the required energy to maintain coronal temperatures. Coronal emission line profiles also indicate minimal wave amplitudes, implying significant energy dissipation before reaching the corona \cite{mein1981mechanical,sakurai2002spectroscopic}. Similarly, while Alfvénic waves are abundant in the solar wind, their contribution to coronal heating appears limited: non-thermal line widths in coronal loops suggest that Alfvén wave energy accounts for less than 25\% of the required heating \cite{hara1999microscopic}. Although turbulence-driven cascades have been observed in the photosphere \cite{petrovay2001turbulence}, chromosphere \cite{reardon2008evidence}, and solar wind \cite{bruno2013solar}, their dissipation mechanisms remain poorly understood. Moreover, additional proposed processes including thermal conduction, viscosity, Landau damping, ion-cyclotron resonance, and stochastic particle acceleration, fail to fully explain the observed heating rates \cite{cranmer2019properties}.

Alternative theories link coronal heating to magnetic activity and solar dynamo processes \cite{sakurai2012helioseismology}, yet significant energy deficits persist. Flares, microflares, and nanoflares are small-scale magnetic reconnection events that follow X-ray power-law distributions but contribute less than 20\% of the corona’s total radiative losses \cite{lin1984solar,parker1988nanoflares,shibata2011solar}. Twisted magnetic structures (e.g., flux ropes, filaments, sigmoids) store nonpotential energy, yet helicity conservation restricts its release \cite{taylor1974relaxation}. Recent direct current models suggest that chromospheric footpoint reconnection could contribute additional heating \cite{priest2018cancellation}, but predictive simulations still fail to match observed thermal structures.

This persistent mismatch between theoretical predictions and observations highlights the need for new dissipation mechanisms. We propose a kinetic-scale process driven by interactions between charged-particle magnetic moments and reconnection-driven field reconfiguration. Unlike conventional wave- or reconnection-based models, this approach directly links macroscopic magnetic topology with particle-scale energy transfer, providing a unified explanation for localized heating and solar wind acceleration.

\section{Proposed Physical Mechanism}
Charged particles in magnetized plasmas execute helical trajectories around magnetic field lines under the Lorentz force, generating a magnetic moment $\mu$ antiparallel to the local field. In equilibrium, $\mu$ remain align with the background field to minimize free energy. Classical theory posits $\mu$ as adiabatically invariant in slowly varying fields:

\begin{equation}
\mu = \frac{\frac{1}{2}m v_\perp^2}{B},
\end{equation}

where $v_\perp$ is the velocity perpendicular to the magnetic field (B). Coupling this to thermal energy via kinetic theory:

\begin{equation}
\frac{1}{2}m v_\perp^2 = \frac{1}{2} K T,
\end{equation}
where $T$ is the particle temperature and $K$ is the Boltzmann constant. It allows $\mu$ estimation from local T and B measurements in the solar atmosphere.

\begin{figure*}
 \includegraphics[width=\textwidth]{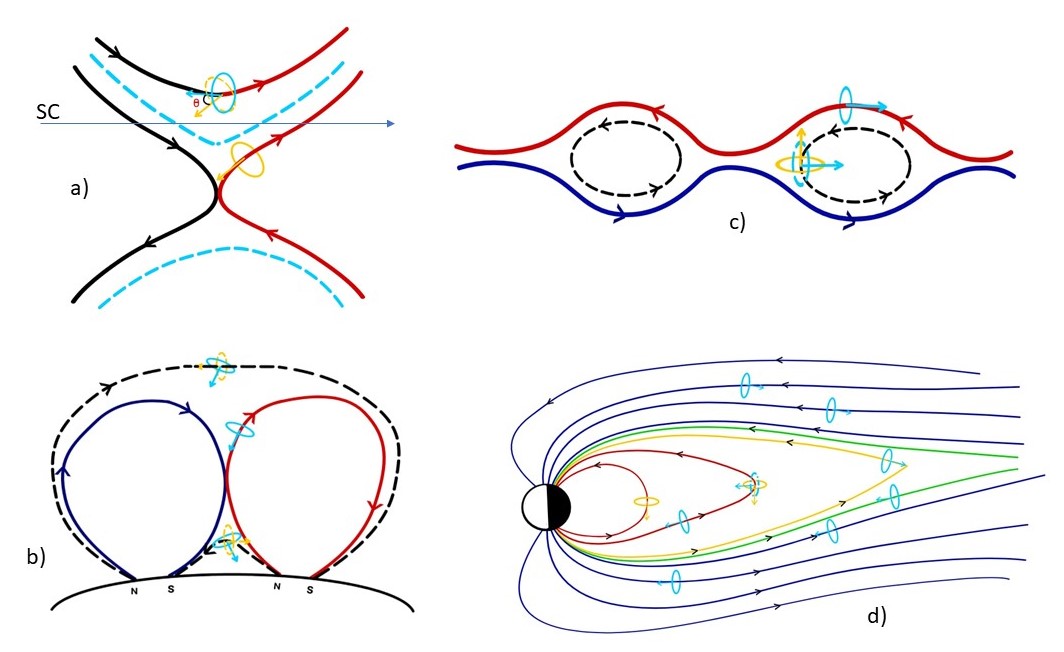}
  \caption{Artistic depiction (not to scale) of magnetic reconnection process in general (top left), and in three environments: solar corona (bottom left), Earth’s magnetotail (bottom right), and a plasmoid (top right). Magnetic field lines  with directional arrows and a representative charged particle’s gyration are shown pre- and post-reconnection. In each image, before reconnection, field lines retain their original geometry, and the particle’s magnetic moment ($\mu$) anti-aligns with the field. After reconnection, field lines reconfigure into nearly perpendicular orientations, thus $\mu$ is misaligned. This drives a restoring torque, forcing the particle toward anti-aligned states. The process enables magnetic to kinetic energy conversion in collisionless plasmas across astrophysical and heliospheric systems.} 
 \end{figure*}

Figure 1 illustrates schematic of magnetic reconnection in different space plasma environment. The figure highlights magnetic field lines reconfiguration during the process.  In regions where the magnetic field changes slowly, charged particles follow well‐defined gyro-motion, and their magnetic moment, $\mu$, is conserved as an adiabatic invariant. However, as particles approach the reconnection X-point, the magnetic field rapidly reorients over very short spatial and temporal scales.  In this critical region, the adiabatic assumption breaks down. The field’s variation becomes too rapid compared to the particle’s gyro-motion, and $\mu$ is no longer conserved.

Prior to the reconnection event, $\mu$ serve as robust adiabatic invariants. This well-defined adiabatic state provides the necessary initial condition for the system. Moreover, at the X-point, the magnetic field configuration changes rapidly, however, the particles retain the memory of their previously conserved magnetic moment. As the field adopts a new configuration, it exerts a torque on the particles. This torque acts to realign the magnetic moment in the anti-parallel direction to the newly oriented field lines.  Consequently,  the work done on the particle as it adjusts to the new field configuration results in an energy gain for the particles, which can be described by the equation:

\begin{equation}
\delta E = \mu \cdot B' = \mu B' \cos \theta,
\end{equation}

Here,  $\delta E$ represents the energy gained by the particle, B' is the magnetic field strength after reconfiguration, and $\theta$ is the angle between $\mu$ and the new field direction. At larger view, the field orientation changes by a maximum of $90^\circ$ in proximity of the reconnection site. Consequently, the magnetic moment of the particles seems to be nearly perpendicular to the newly oriented field lines.  If the magnetic field strength stays nearly constant, the particle can gain energy up to its initial amount, implying up to $100\%$ energy increase. Variations in B will modulate this gain, with a stronger field enhancing the energy transfer and a weaker field reducing it.  

Thus, the effect is implemented by using the initial adiabatic state as a baseline and explains how the sudden loss of adiabatic condition at the X-point results in energy gain and chaotic particle trajectories. Historical studies, such as those by Speiser (1965) \cite{speiser1965particle}, have shown that in regions where the magnetic field reverses, particle paths become chaotic. More recent research on kinetic reconnection \cite{drake2006electron, hoshino2012stochastic} supports this, showing that particles entering the diffusion region become unmagnetized and are accelerated by the reconnection electric fields.
However, it remains unclear whether the attached particles respond instantaneously or with a finite response time to this sudden reorientation. Understanding this response time remains an open question for future investigations.

In summary, from the particle’s perspective, a rapidly changing magnetic field generates an electric field that accelerates the particle. In the field frame,field frame, sudden loss of adiabatic invariance causes a sharp increase in the particle’s kinetic energy/ temperature, altering its magnetic moment. This change leads to chaotic motion, detaching the particle from its field lines. As a result, some particles are expelled as exhaust, while others remain attached with increased energy and a modified magnetic moment. We term this interaction between the magnetic field and the particle’s magnetic moment as the Raghav effect, which has significant implications for particle energization in astrophysical and space plasma environments.

\section{Observation}
A proof of concept is required to establish the physical mechanism. The most direct way to test this hypothesis is through in situ measurements at a reconnection site. We analyze the passage of ambient solar wind through the Wind spacecraft using 3-second resolution data from the Magnetic Field Investigation (MFI) and 3D Plasma and Energetic Particle (3DP) instruments onboard Wind. We examined an event on October 13, 2002, from 22:00:00 to 22:15:00 UT, during which interplanetary conditions (Fig. 2) indicate a reconnection site crossing. The top panel of Fig. 2 shows the interplanetary magnetic field (IMF) strength (\textbf{B}), with the event highlighted in cyan. During this interval, \textbf{B} decreases sharply from $\sim 6$ nT to a minimum of 2 nT, while all components of the IMF (Bx, By, Bz) exhibit coherent variations. The trajectory of the spacecraft, marked by the blue arrow in Fig. 1(a), suggests a possible passage through a reconnection site. It is supported by concurrent increases in plasma density, temperature, bulk velocity, and plasma beta.

During the interval marked by the cyan-shaded region, the plasma temperature rises sharply from a pre-event minimum of 58,262 K to 101,522 K, an increase of 43,260 K ($73.6\%$), as the spacecraft traverses the reconnection site. This is below the theoretical limit of energy transfer of 100\%, indicating that the actual transfer depends on cos($\theta$) value and the spacecraft’s proximity to the reconnection outflow. While these findings validate the hypothesis, statistical studies of multiple reconnection site crossings are needed to generalize the proposed mechanism.

 \begin{figure*}
 \includegraphics[width=\textwidth]{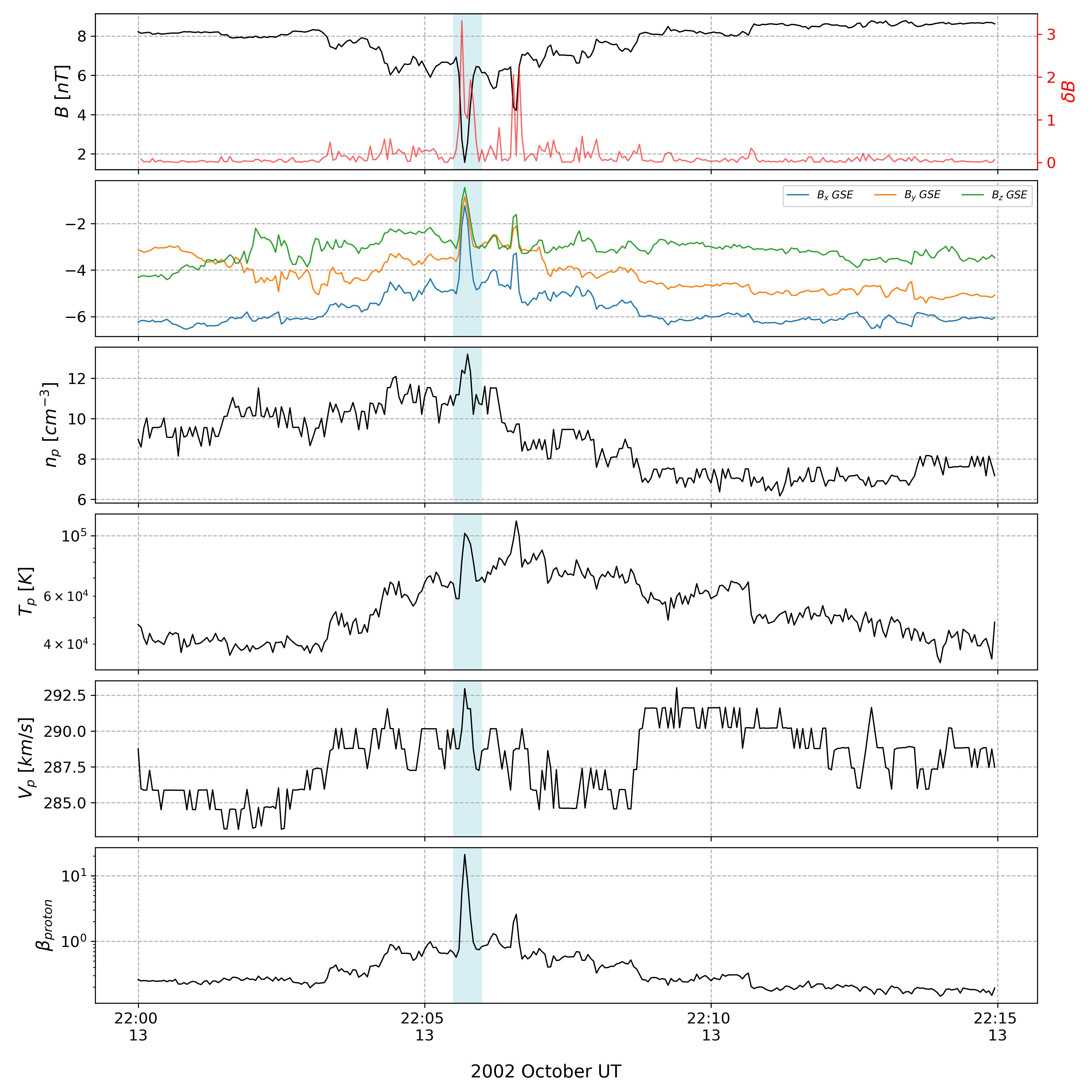}
  \caption{Interplanetary parameters of the studied event. These are observations by Wind on October 13 2002 from 22:00 UT to 22:15 UT. The top-to-bottom panels represent magnetic field strength (\textbf{B}) and its fluctuations, magnetic field components, proton number density,  proton temperature, solar wind proton velocity, and plasma beta. The cyan-shaded region marks the area of interest.  } 
 \end{figure*}

\section{Implication}

Observations of the solar photosphere reveal a highly structured, clumpy magnetic field concentrated into small-scale elemental flux tubes that extend into the corona \cite{solanki1993small, muller1994properties, stenflo1998differential}. Turbulent convection continuously drives the evolution of these flux tubes, causing translational and rotational motions that tangle and twist the magnetic field lines. Parker (1972) \cite{parker1972topological} suggested that these slow footpoint motions progressively stress the coronal magnetic field, resulting in tangled, twisted, and braided configurations that eventually reconnect to release stored energy. In a mixed-polarity corona, magnetic reconnection involves interactions among open-open (bipolar), open-closed (tripolar), or closed-closed (quadrupolar) field configurations, continuously reshaping the coronal magnetic topology.

The Sun’s photospheric temperature is approximately $5,778 K$ ($\sim 5,800 K$). According to the proposed Raghav effect, plasma particles can gain energy during successive reconnection events. Under the assumption of 100\% energy transfer efficiency, each reconnection event essentially doubles a particle’s temperature—raising it to about $1.16 \times 10^4$ K after the first event, $2.32 \times 10^4$ K after the second, and $4.64 \times 10^4$ K after the third. By the eighth reconnection cycle, the particle’s temperature can reach approximately $1.1 \times 10^6$ K, in agreement with observed coronal temperatures. Even with a lower efficiency, only a modest increase in reconnection events is required to achieve coronal heating.  Furthermore, if a plasma particle accumulates sufficient energy through multiple reconnection events, it can exceed the solar escape velocity and contribute to the formation of the solar wind.  This framework offers a unified explanation for coronal heating and solar wind acceleration, indicating that reconnection-driven particle energization via the Raghav effect is crucial in solar and astrophysical plasma dynamics.

The Raghav effect may represent a universal mechanism for particle energization in interstellar and astrophysical plasmas. By facilitating efficient energy transfer during rapid magnetic reconnection, this effect could explain the acceleration of particles observed in environments ranging from stellar coronae and magnetospheres to accretion disks and interstellar clouds. In these diverse settings, the interplay between turbulent magnetic fields and charged particles is critical for driving phenomena such as cosmic ray acceleration, plasma heating, and the onset of explosive events like flares or gamma-ray bursts. Recognizing the Raghav effect as a fundamental process not only enriches our understanding of energy dynamics in high-energy astrophysical systems but also provides a unifying framework to study how localized magnetic interactions can influence large-scale plasma behavior across the universe.

\section{ACKNOWLEDGEMENT}

The author gratefully acknowledges Mr. Shubham Kadam for converting the original pen-and-paper sketch into the digital illustration shown in Figure 1. Additionally, sincere thanks are extended to Mr. Kalpesh Ghag, Ms. Marium Karari, Mr. Ajay Kumar, and Ms. Urvi Naik for their discussions.

\nocite{*}

\bibliography{Corona}

\end{document}